# 3 PORTFOLIO MANAGEMENT APPROACH IN TRADE CREDIT DECISION MAKING

**Grzegorz MICHALSKI**[*]

## Abstract

The basic financial purpose of an enterprise is maximization of its value. Trade credit management should also contribute to realization of this fundamental aim. Many of the current asset management models that are found in financial management literature assume book profit maximization as the basic financial purpose. These book profit-based models could be lacking in what relates to another aim (i.e., maximization of enterprise value). The enterprise value maximization strategy is executed with a focus on risk and uncertainty. This article presents the consequences that can result from operating risk that is related to purchasers using payment postponement for goods and/or services. The present article offers a method that uses portfolio management theory to determine the level of accounts receivable in a firm. An increase in the level of accounts receivables in a firm increases both net working capital and the costs of holding and managing accounts receivables. Both of these decrease the value of the firm, but a liberal policy in accounts receivable coupled with the portfolio management approach could increase the value. Efforts to assign ways to manage these risks were also undertaken; among them, special attention was paid to adapting assumptions from portfolio theory as well as gauging the potential effect on the firm value.

## 1. Introduction

The basic financial aim of an enterprise is maximization of its value. At the same time, both theoretical and practical meaning is researched for determinants that increase the enterprise value. Financial literature contains information about numerous factors that influence enterprise value. Among those contributing factors is the extent of the net working capital and the elements shaping it, such as the level of cash tied up in accounts receivable, inventories, the early settlement of accounts payable, and operational cash balances. The greater part of classic financial models and proposals relating to optimum current assets management was constructed with net profit maximization in mind. This is the reason why these models need reconstruction in order to make them suitable to firms that want to maximize their value. The decision if extend the trade credit terms, is a compromise between limiting the risk of allowing for the payment postponement from unreliable purchasers and gaining new customers by way of a more liberal enterprise trade credit policy. This decision shapes the level and quality of accounts receivable.

---
[*] *Dr. Grzegorz Michalski, Wroclaw University of Economics, Department of Corporate Finance and Value Management, ul. Komandorska 118/120, pok. 704-Z, PL53-345 Wroclaw, Poland, Grzegorz.Michalski@ae.wroc.pl; http://michalskig.ae.wroc.pl/*

The question discussed in this article concerns the possibility of using portfolio theory in making decisions about selecting which customers should be given trade credit. In this article, we will show that it is possible that the firm can sell on trade credit terms to some customers, who previously were rejected because of too great an operational risk, with a positive outcome the creation of increased firm value. This extension of trade credit is possible only if the firm has purchasers from various branches, and if these branches have different levels of operating risk. The key to success for a firm is to perform portfolio analysis with the result of a varied portfolio of customers with a spectrum of managed levels of operating risk.

## 2. Value Based Management of Accounts Receivable

If holding accounts receivable on a level defined by the enterprise provides greater advantages than negative influence, the firm value will grow. Changes in the level of accounts receivable affect on the value of the firm. To measure the effects that these changes produce, we use the following formula, which is based on the assumption that the firm present value is the sum of the future free cash flows to the firm ($FCFF$), discounted by the rate of the cost of capital financing the firm:

$$\Delta V_p = \sum_{t=1}^{n} \frac{\Delta FCFF_t}{(1+k)^t},$$ (1)

where $\Delta V_p$ = firm value growth; $\Delta FCFF_t$ = future free cash flow growth in period $t$; and $k$ = discount rate[1].

Future free cash flow is expressed as:

$$FCFF_t = (CR_t - CE_t - NCE) \times (1-T) + NCE - Capex - \Delta NWC_t$$ (2)

where $CR_t$ = cash revenues on sales; $CE_t$ = cash expenses resulting from fixed and variable costs in time $t$; $NCE$ = non-cash expenses; $T$ = effective tax rate; $\Delta NWC$ = net working growth; and $Capex$ = capital expenditure resulting from the growth of operational investments (money used by a firm to acquire or upgrade physical assets, such as property, industrial buildings, or equipment).

---

[1] *To estimate changes in accounts receivable levels, we accept discount rate equal to the average weighted cost of capital (WACC). Such changes and their results are strategic and long term in their character, although they refer to accounts receivable and short run area decisions (T.S. Maness 1998, s. 62-63).*

Similar conclusions, related to the results of changes in trade credit policy on the firm value, can be estimated on the basis of economic value added, the extent to which residual profit (the added value) increased the value of the firm during the period:

$$EVA = NOPAT - k \times (NWC + OI),$$ (3)

where $EVA$ = economic value added; $NWC$ = net working capital; $OI$ = operating investments; and $NOPAT$ = net operating profit after tax, estimated on the basis of the formula:

$$NOPAT = (CR_t - CE_t - NCE) \times (1 - T)$$ (4)

The net working capital (NWC) is the part of current assets that is financed with fixed capital. The net working capital (current assets less current liabilities) results from lack of synchronization of the formal rising receipts and the real cash receipts from each sale. It is also caused by a divergence during time of rising costs and time when a firm pays its accounts payable.

$$NWC = CA - CL = AAR + INV + G - AAP$$ (5)

where $NWC$ = net working capital; $CA$ = current assets; $CL$ = current liabilities; $AAR$ = average level of accounts receivable; $INV$ = inventory; $G$ = cash and cash equivalents; and $AAP$ = average level of accounts payable.

During estimation of the free cash flows, the holding and increasing of net working capital ties up money used for financing net working capital. If net working capital increases, the firm must utilize and tie up more money, and this decreases free cash flows. Production level growth necessitates increased levels of cash, inventories, and accounts receivable. Part of this growth will be covered with current liabilities that automatically grow with the growth of production and sales. The remaining cash requirements (that are noted as net working capital growth, $\Delta NWC$) will require a different form of financing.

Trade credit policy decisions changing the terms of trade credit create a new accounts receivable level. Consequently, trade credit policy has an influence on firm value. This comes as a result of alternative costs of money tied in accounts receivable and general costs associated with managing accounts receivable. Both the first and the second involve modification of future free cash flows and as a consequence firm value changes. In Figure 1, we show the influence of trade credit policy changes on firm value. These decisions change:

- future free cash flows ($FCFF$),
- life of the firm ($t$) and
- rate of the cost of capital financing the firm ($k$).

Changes to these three components influence the creation of the firm value ($\Delta Vp$).

Accounts receivable changes (resulting from changes in trade credit policy of the firm) affect the net working capital level and also the level of accounts receivable management operating costs in a firm; these operating costs are a result of accounts receivable level monitoring and recovery charges).

## The trade credit policy influence on firm value

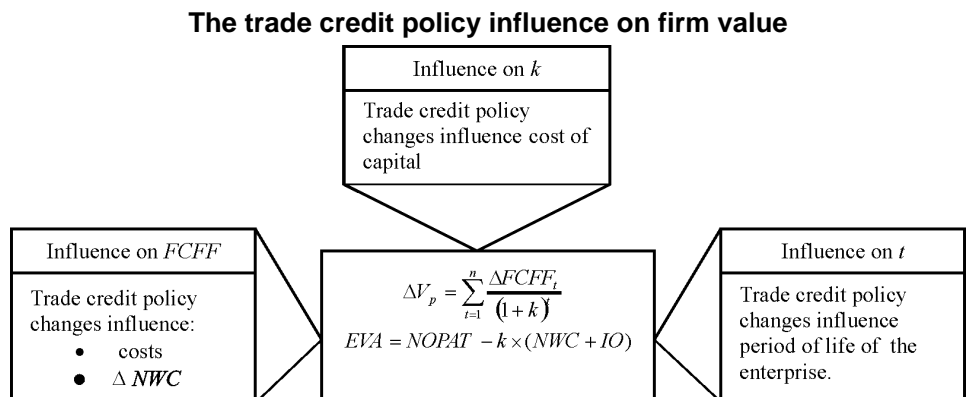

where: *FCFF* = free cash flows to firm; Δ*NWC* = net working capital growth; *k* = cost of the capital financing the firm; and *t* = the lifetime of the firm and time to generate single *FCFF*.

Source: own study.

Trade credit terms give evidence about a firm trade credit policy. They are the parameters of trade credit and include:

- the maximum delay in payment by purchasers (trade credit period);
- the time the purchaser has to pay with a cash discount; and
- the rate of the cash discount.

The length of the cash discount period and the maximum delay in payment by purchasers give information about the character of a firm trade credit policy. These trade credit conditions are:

$$ps/os, \text{net } ok \tag{6}$$

where *ps* = cash discount rate, *os* = cash discount period, and *ok* = maximum payment delay period.

The terms of a trade credit sale are the result of a firm's management decision made on the basis of information about factors such as:

- market competition,
- the kind of goods or services offered,
- seasonality and elasticity of demand,
- price,
- type of customer, and
- profit margin from sale.

It is important to match the length of the trade credit of a firm to its customer's capabilities. The enterprise giving the trade credit should take into account the purchasers' inventory conversion cycle as well as its accounts receivable conversion cycle. These two elements make up the operating cycle of a purchaser. The shorter this cycle, the shorter should be the maximum payment delay period offered to a purchaser. The maximum payment delay period for purchaser is the maximum expected period of accounts receivable cycle for a seller.

In order to choose what terms of sale should be proposed to the purchaser a firm management can use the incremental analysis as final criterion, as well as compare the influence of these proposals on firm value. Incremental analysis is a tool for estimating the effects of changes in trade credit policy on the enterprise. This analysis usually takes into account three basic elements:

(1) Estimation of the results of changes on sales as well as losses resulting from bad debts.

(2) Estimation of the changes in the firm accounts receivable level.

Accounts receivable growth we have as:

$$\Delta AAR = \left(ACP_1 - ACP_0\right) \times \frac{CR_0}{360} + VC \times ACP_1 \times \frac{CR_1 - CR_0}{360}, \text{ if } CR_1 > CR_0$$

$$\Delta AAR = \left(ACP_1 - ACP_0\right) \times \frac{CR_1}{360} + VC \times ACP_0 \times \frac{CR_1 - CR_0}{360}, \text{ if } CR_1 \le CR_0$$

(7)

where $\Delta AAR$ = accounts receivable growth; $ACP_0$ = receivables collection period before trade credit policy change; $ACP_1$ = receivables collection period after trade credit policy change; $CR_0$ = cash revenue before trade credit policy change; $CR_1$ = cash revenue after trade credit policy change; and $VC$ = variable costs (in percent from sales incomes).

(3) Estimation of the firm value change:

$$\Delta EBIT = [(CR_1 - CR_0) \times (1 - VC) - k_{AAR} \times \Delta AAR +$$
$$- (l_1 \times CR_1 - l_0 \times CR_0) - (sp_1 \times CR_1 \times w_1 - sp_0 \times CR_0 \times w_0)]$$

(8)

where $\Delta EBIT$ = earnings before interests and taxes growth; $k_{AAR}$ = operating costs of accounts receivable management in a firm; $l_0$ = bad debts loses before trade credit policy change; $l_1$ = bad debts loses after trade credit policy change; $sp_0$ = cash discount before trade credit policy change; $sp_1$ = cash discount after trade credit policy change; $w_0$ = part of purchasers using cash discount before trade credit policy change; and $w_1$ = part of purchasers using cash discount after trade credit policy change.

To check how changes in the accounts receivable level and *EBIT* influence on firm value, it is possible to use changes in future free cash flows. First, we have changes in *FCFF* in time 0:

$$\Delta FCFF_0 = -\Delta NWC = -\Delta AAR$$

(9)

Next, the free cash flows to firm in periods (from 1 to *n*), as:

$$\Delta FCFF_{1...n} = \Delta NOPAT = \Delta EBIT \times \left(1 - T\right)$$

(10)

**Example 1.** An enterprise $CR_0$ = 500 000 000 €. $VC$ = 50% × $CR$. Operating costs of accounts receivable management in a firm, $k_{AAR}$ = 20%. WACC = 15%. T = 19%. Before trade credit policy change half of firm customers pay before delivery. 25% of them use 2% cash discount paying on the $10^{th}$ day. The remaining customers pay on the $30^{th}$ day. Bad debts losses 3% of $CR$. The trade credit policy changes (from 2/10, net 30 to 3/10, net 40) considered by firm will result: 40% of firm customers will pay before delivery. 30% of them use 3% cash discount paying on the $10^{th}$ day. The remaining customers pay on the $45^{th}$ day. Bad debts losses 4% of $CR$. New $CR_1$ = 625 000 000 €. The effects of changes in trade credit policy would be felt for 3 years.

Because 50% of sale before change of policy, is done in cash, 25% on principle of collected on the $30^{th}$ day, 25% on principle of charge regulated up to the $10^{th}$ day, then the $ACP_0$ is:

$$ACP_0 = 0,5 \times 0 + 0,25 \times 10 + 0,25 \times 30 = 10 \text{ days.}$$

The $ACP_1$ after change is:

$$ACP_1 = 0,4 \times 0 + 0,3 \times 10 + 0,3 \times 45 = 16,5 \text{ days.}$$

That is why expected increase of average level of accounts receivable will be:

$$\Delta AAR = \left(16,5 - 10\right) \times \frac{500000000}{360} + 0,5 \times 16,5 \times \frac{125000000}{360} = 11\,892\,361 €.$$

Therefore in result of trade credit policy change, the average state of accounts receivable will grow up for 11 892 361 €.

Next, we have $\Delta EBIT$:

$$\Delta EBIT = 125000000 \times 0,5 - 20\% \times 11892361 -$$
$$- (4\% \times 625000000 - 3\% \times 500000000) +$$
$$- (3\% \times 625000000 \times 30\% - 2\% \times 500000000 \times 25\% = 46\,996\,527,8 €$$

Using equations nine and ten, we can estimate firm value growth:

$$\Delta V = -11\,892\,361 + \frac{46\,996\,527,8 \times 0,81}{0,15} \times \left(1 - \frac{1}{1,15^3}\right) = 75\,023\,598 €.$$

As we see, the trade credit policy change will increase the firm value. Similar information is given by estimation of Δ$EVA$ after trade policy change:

$$\Delta EVA = 0,81 \times 46\,996\,527,8 - 15\% \times 11\,892\,361 = 36\,283\,333 €.$$

As one can see through the discussed case, the first half and then 40% of sales are realized on the principle of cash sale. This results from the fact that those customers who created sales only for cash did not fulfill the requirements relating the risks, which is considered as percentage of delayed payments. Therefore, the firm stopped offering these purchasers sales on the principle of the trade credit. This was despite the fact that their financing with the trade credit made it possible to notice much greater activity and larger level of income from sales, than at the trade credit relinquishment.

## 3. Portfolio Theory Approach in Trade Credit Decisions

A portfolio is a set of assets (for example, accounts receivable). The portfolio approach to accounts receivable management can be used by utilizing the rate of profit (rate of advantage from assets) as one of the basic criteria that the firm giving the trade credit should encourage the purchaser to consider when making decisions (Jajuga 1994, p. 80-110). The profit rate resulting from the trade credit can be defined as:

$$R_{nAR} = \frac{\Delta CR - \Delta Costs}{\Delta Costs} \qquad (11)$$

where $R_{nAR}$ = profit rate from giving the trade credit to purchasers $n$, $\Delta CR$ = cash revenue growth generated from additional sale to $n$ customers instead of the cash sale, and $\Delta Costs$ = growth of costs resulting from offering the trade credit to purchaser $n$.

The present rate of profit is realized amid conditions of risk and uncertainty. The rate of profit changes varies according to the various probabilities. These probabilities result from customers' marketable situations which influence their ability to regulate their accounts payable to the seller in an appropriate manner. The risk measure connected with the accounts receivable of a concrete purchaser varies according to the following equation:

$$V = \sum_{i-1}^{m} p_i \times \left(R_i - R\right)^2 \qquad (12)$$

where $p_i$ = based on historical data probability of $R_i$, and $R_i$ = expected rate of return from accounts receivable from the group of purchasers $i$.

The measure of risk also can be defined according to standard deviation:

$$s = \sqrt{V} = \sqrt{\sum_{i-1}^{m} p_i \times \left(R_i - R\right)^2} \qquad (13)$$

Both the variation and the standard deviation can be estimated for the historical data of a purchaser.

The next element is the correlation of profit from the trade credit given to the purchaser (or to the group of purchasers) in which the profits of the trade credit are given to other purchasers (or to different groups of purchasers). If the firm completes the transactions with more than one group of purchasers, it is possible to distinguish two or more homogeneous groups in relation to the risk and profit from giving the trade credit. In this case, the portfolio approach can be used. These groups can belong to definite trades[1], and a connection does or can exist between the accounts receivables of these groups of purchasers. The measure of such a connection is usually a coefficient of correlation:

$$\rho_{1.2} = \frac{\sum_{i=1}^{m} p_i \times (R_{1i} - R_1) \times (R_{2i} - R_2)}{s_1 \times s_2}$$

(14)

where $\rho_{1.2} =$ coefficient of first and second group of accounts receivable correlation; $R_1 =$ expected rate of return from accounts receivable of the first group of purchasers; $R_2 =$ expected rate of return from accounts receivable of the second group of purchasers; $s_1 =$ standard deviation for the first group; $s_2 =$ standard deviation for the second group; $R_{1i} =$ individual rate of return from accounts receivable of purchaser $i$ from the first group of purchasers; $R_{2i} =$ individual rate of return from accounts receivable of purchaser $i$ from the second group of purchasers; and $p_i =$ probability of individual rate of return from accounts receivable of purchaser $i$.

---

[1] *In Polish business practice purchasers coming from one trade group have similar payment because they serve the same market and have similar customers with similar payment habits.*

To show how portfolio approach can be used in accounts receivable management, we will use the portfolio of two groups of accounts receivable as example.

**Example 1.** The firm cooperates with two homogenous groups of purchasers. The first group of purchasers delivers its services to industry A, the second group of purchasers serves customers from industry B. Creating a portfolio of two kinds of accounts receivable makes sense only when the correlation between profits from the giving trade credit for these groups is less than 1.

**Example 2, Case 1.** Correlation coefficient between accounts receivable profits from Groups A and B equals 1, $\rho_{A,B} = 1$.

**Figure 2**

**The profit - risk relation for portfolio of accounts receivable for two groups of purchasers if $\rho_{A,B} = 1$.**

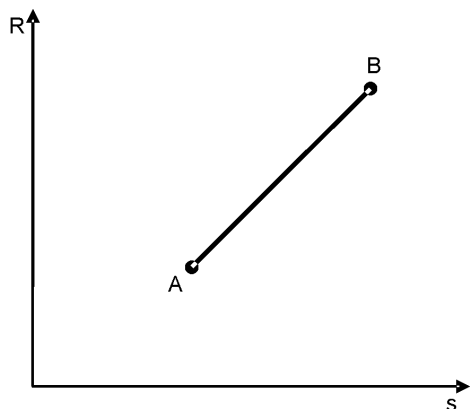

Source: Own study, on the basis of Jajuga, 1994.

Figure 2 shows that there is no possibility of increasing profit from diversification without increasing risk if $\rho_{A,B} = 1$.

**Example 2, Case 2.** Coefficient of correlation equal (- 1), $\rho_{A,B} = (-1)$. Perfect negative correlation.

**Figure 3**

**The profit - risk relation for portfolio of accounts receivable for two groups of purchasers if $\rho_{A,B} = (-1)$.**

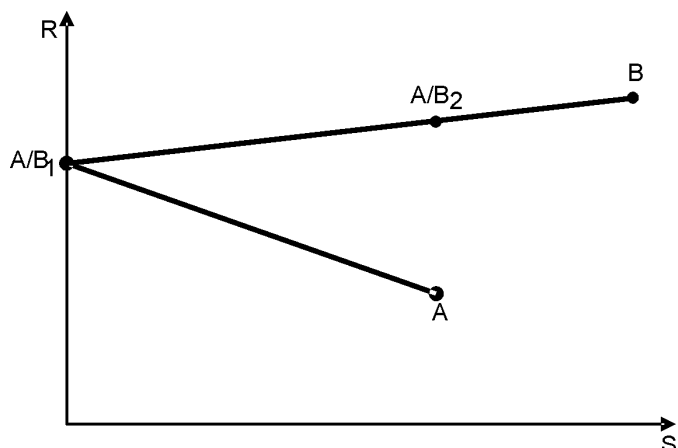

Source: Own study, on the basis of Jajuga, 1994.

At Point A, we offer trade credit only to Group A. At point B we offer trade credit to Group B. If we are following from Point A (and we are enlarging contribution of Group B in accounts receivable portfolio) to A/B$_1$, the risk *s* is decreasing and the profit *R* is increasing. As we see in Figure 3, it makes no sense to possess accounts receivable only from Group A. It is because with identical risk *s*, the portfolio A/B$_2$ offers higher profit *R*.

**Example 2, Case 3.** Coefficient of correlation equal 0, $\rho_{A,B} = 0$. It is a situation when benefits from giving trade credit to Group A and Group B are not related to each other in any way.

In such situation, the only possible is partial reduction of risk. The reasonable firm should not choose any portfolio of charge lying on the A - A/B$_3$ line, because it is always possible to find more profitable equivalent on: A/B$_3$ - A/B$_4$ line, which with the same risk *s* gives higher profit *R*. The skilful construction of two groups of accounts receivable portfolio can lead to considerable reduction of risk. The inclusion to single-asset portfolio second component, almost always leads to risk decreasing, sometimes even with simultaneous profit growth (Brigham 2004, p. 77; Jajuga 1994, p. 119; Jajuga 1997; Jajuga 1993; Wait 2002; Fabozzi 2000; Jajuga 2002).

**The profit - risk relation for portfolio of accounts receivable for two groups of purchasers if $\rho_{A,B} = 0$.**

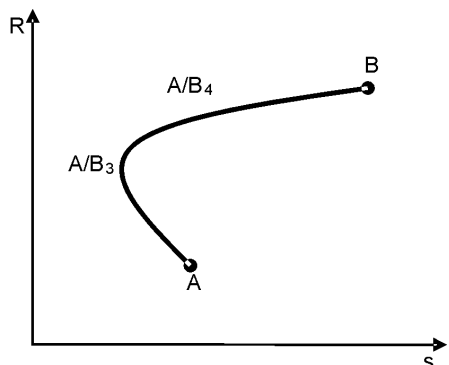

Source: Own study, on the basis of Jajuga, 1994.

**Example 3** (continuation of example 1). After the historical data analysis had been achieved, firm managers noticed that the expected profits were higher and correlated negatively with profits generated from purchases by current customers. This was certainly from allowing trade credit to customers who had made cash purchases because of the high risk during the receivables collection period. These trends lead to the expectation[1] of a lower risk of profits from accounts receivable and growth in profits from sales in general at the same time. A 3% cash discount was proposed for customers who paid within 10 days along with an extension of the payment deadline to 45 days for any remaining customers. As a result, 4% of sales would be paid for with in cash while 40% of customers would take advantage of the cash discount by paying by the 10th day. Remaining customers (46% of sales) would make their payments on the 45th day. Bad debts = 1% × *CR*. $CR_1$ = 700 000 000 €. The effect of these changes in trade credit policy would be felt for three years. In addition, *VC* would be reduced from 50% to 49% thanks to the positive advantages of scales resulting from larger sales (and increased production).

So, we have:

$$DSO_2 = 0{,}04 \times 0 + 0{,}40 \times 10 + 0{,}46 \times 45 = 24{,}7 \text{ days}$$

$$\triangle AAR = \left(24{,}7 - 10\right) \times \frac{500\,000\,000}{360} + 0{,}49 \times 24{,}7 \times \frac{200\,000\,000}{360} = 27\,140\,556 \text{ €.}$$

$$\triangle EBIT = 200\,000\,000 \times 0{,}51 - 20\% \times 27\,140\,556 - \left(1\% \times 700\,000\,000 - 3\% \times 500\,000\,000\right) +$$
$$- \left(3\% \times 700\,000\,000 \times 40\% - 2\% \times 500\,000\,000 \times 25\% = 98\,671\,889 \text{ €}\right.$$

---

[1] With admittance of taking both groups simultaneously on trade credit principles.

From this, we have the following change in the firm value:

$$\Delta V = -27\ 140\ 556 + \frac{98\ 671\ 889 \times 0{,}81}{0{,}15} \times \left(1 - \frac{1}{1{,}15^3}\right) = 155\ 344\ 454\ €.$$

The firm value will increase. The proposed change will be more profitable than in example 1 without using the portfolio approach. This related information comes from the estimation of *EVA* growth:

$$\Delta EVA = 0{,}81 \times 98\ 671\ 889 - 15\% \times 27\ 140\ 556 = 75\ 853\ 147\ €.$$

# 4. Conclusions

Accounts receivable management decisions are very complex. On the one hand, too much money is tied up in accounts receivables, because of an extreme liberal policy of giving trade credit. This burdens the business with higher costs of accounts receivable service with additional high alternative costs. Additional costs are further generated by bad debts from risky customers. On the other hand, the liberal trade credit policy could help enlarge income from sales. In the article, the problem was linked to the operational risk of purchasers interested in receiving trade credit who, as separately considered groups, may characterize too high risk level. However, if they are considered as one of several groups of enterprise customers, and if their payment habits are correlated with the payment habits of the remaining groups, what was formerly impossible could become possible, and may even turn profitable. The portfolio of assets, like the portfolio of accounts receivables, sometimes presents a lower risk to acceptable advantages than the independently considered groups of purchasers.